\documentclass[prd,preprint,superscriptaddress,showpacs,byrevtex]{revtex4}
\usepackage{bm}
\def\fsl#1{\setbox0=\hbox{$#1$}           
   \dimen0=\wd0                                 
   \setbox1=\hbox{/} \dimen1=\wd1               
   \ifdim\dimen0>\dimen1                        
      \rlap{\hbox to \dimen0{\hfil/\hfil}}      
      #1                                        
   \else                                        
      \rlap{\hbox to \dimen1{\hfil$#1$\hfil}}   
      /                                         
   \fi}                                         %
\newcommand{\be}{\begin{equation}}
\newcommand{\ee}{\end{equation}}
\newcommand{\bea}{\begin{eqnarray}}
\newcommand{\eea}{\end{eqnarray}}
\newcommand{\beq}{\begin{equation}}
\newcommand{\eeq}{\end{equation}}
\newcommand{\beqs}{\begin{eqnarray}}
\newcommand{\eeqs}{\end{eqnarray}}

\newcommand{\dslash}{D\hspace{-0.067in}\slash}
\begin{document}
\title{ Gauge Fixing Identity in the Background Field Method of QCD in Pure Gauge }
\author{Gouranga C Nayak } \email{nayak@physics.arizona.edu}
\affiliation{ Department of Physics, University of Arizona, Tucson, AZ 85721, USA}
\begin{abstract}
In this paper we derive a gauge fixing identity by varying the covariant gauge fixing term in
$Z[A,J,\eta, {\bar \eta}]$ in the background field method of QCD in pure gauge. Using this gauge
fixing identity we establish a relation between $Z[J,\eta,{\bar \eta}]$ in QCD and $Z[A,J,\eta, {\bar \eta}]$ 
in background field method of QCD in pure gauge. We show the validity of this gauge fixing identity in general 
non-covariant and general Coulomb gauge fixings respectively. This gauge fixing identity is used to prove factorization 
theorem in QCD at high energy colliders and in non-equilibrium QCD at high energy heavy-ion colliders. 
\end{abstract}
\maketitle
\pagestyle{plain}
\pagenumbering{arabic}
\section{Introduction}
Background field method of QCD was originally formulated by 't Hooft \cite{thooft} and later
extended by Abbott \cite{abbott}. This is an elegant formalism which can be useful to construct gauge invariant
(off-shell) green's functions in QCD. For example, certain properties of structure functions and/or fragmentation functions at
high energy colliders and at high energy heavy-ion colliders may be studied by using these gauge invariant green's
functions. This formalism is also useful to study quark and gluon production from classical chromo field \cite{peter}
via Schwinger mechanism \cite{schw}, to compute $\beta$ function in QCD \cite{peskin} and to study evolution of QCD
coupling constant in the presence of chromofield \cite{nayak}.

Unlike QCD, since the lagrangian density in the generating functional in the background field method of QCD is
gauge invariant (even after quantizing the theory), this formalism may be useful to study properties of
certain gauge invariant non-perturbative physical quantities in QCD. From
this point of view it is desirable to find situations where we can relate generating functional
in the background field method of QCD to the generating functional in QCD ({\it i.e.} QCD without the
background field). In the presence of external sources a relation between the generating functional in QCD and
the generating functional in the background field method of QCD in pure gauge
\bea
T^aA_\mu^a (x)= \frac{1}{ig}(\partial_\mu U) ~U^{-1},~~~~~~~~~~~~~U=e^{igT^a\beta^a(x)}
\label{gtqcd}
\eea
may be useful in many physical situations.

However, unlike QED \cite{tucci}, finding an exact relation between the generating
functional $Z[J,\eta,{\bar \eta}]$ in QCD and the generating functional $Z[A,J,\eta,{\bar \eta}]$
in the background field method of QCD in pure gauge is not easy. The main difficulty is due to the
gauge fixing terms which are different in both the cases. While the Lorentz (covariant) gauge fixing
term in QCD is independent of the background field $A_\mu^a$, the background field gauge fixing
term in the background field method of QCD depends on $A_\mu^a$ \cite{thooft,abbott}.
Hence it might be useful to obtain a gauge fixing identity by varying the gauge fixing term in $Z[A,J,\eta, {\bar \eta}]$
in the background field method of QCD in order to find a relation between
$Z[J,\eta, {\bar \eta}]$ in QCD and $Z[A,J,\eta, {\bar \eta}]$ in background field method of QCD in pure gauge.
In QCD without background field, a general gauge fixing WT identity was obtained in \cite{justin}.

In this paper we will derive a gauge fixing identity by varying the covariant gauge fixing term in $Z[A,J,\eta,{\bar \eta}]$
in the background field method of QCD in pure gauge. We will make a detailed analysis of this identity by using general non-covariant and
general Coulomb gauge fixing terms. We will show that the gauge fixing identity holds for covariant,
general non-covariant and general Coulomb gauge fixings respectively. Using this gauge fixing identity we will establish a relation between
$Z[J,\eta,{\bar \eta}]$ in QCD and $Z[A,J,\eta, {\bar \eta}]$ in background field method of QCD in pure gauge.

We have used this gauge fixing identity in \cite{nkv} to prove factorization of soft and collinear divergences at high energy
colliders. We have also used this identity in \cite{nkm} to prove factorization of fragmentation function in non-equilibrium
QCD which can be experimentally applicable at high energy heavy-ion colliders at RHIC and LHC \cite{nayakfr}.

The paper is organized as follows. We derive a gauge fixing identity
by varying the covariant gauge fixing term in $Z[A,J,\eta, {\bar \eta}]$ in background field
method of QCD in pure gauge in section II. In section III and IV we perform our calculation
by using general non-covariant and general Coulomb gauge fixings respectively. We establish
a relation between $Z[J,\eta,{\bar \eta}]$ in QCD and $Z[A,J,\eta, {\bar \eta}]$ in background
field method of QCD in pure gauge by using this gauge fixing identity in section V. Section
VI contains conclusions.
\section{ Derivation of Gauge Fixing Identity by Using Covariant Gauge Fixing }
In the background field method of QCD the generating functional is given by \cite{thooft,abbott}
\bea
&& Z^{G}[A,J,\eta,{\bar \eta}]=\int [dQ] [d{\bar \psi}] [d \psi ] ~{\rm det}(\frac{\delta G^a(Q)}{\delta \omega^b}) \nonumber \\
&& e^{i\int d^4x [-\frac{1}{4}{F^a}_{\mu \nu}^2[A+Q] -\frac{1}{2 \alpha}
(G^a(Q))^2+{\bar \psi} \dslash [A+Q] \psi + J \cdot Q +{\bar \eta} \psi + \eta {\bar \psi} ]}
\label{zaqcd}
\eea
where the covariant gauge fixing term is
\bea
G^a(Q) =\partial_\mu Q^{\mu a} + gf^{abc} A_\mu^b Q^{\mu c}=D_\mu[A]Q^{\mu a}
\label{ga}
\eea
which depends on the background field $A_\mu^a$. Under the infinitesimal gauge transformation
\bea
\beta << 1,
\label{b1}
\eea
the gluon field $Q_\mu^a$ and the background field $A_\mu^a$ transform as follows \cite{abbott}
\bea
&& \delta Q_\mu^a = -gf^{abc}\beta^b (A_\mu^c +Q_\mu^c) + \partial_\mu \beta^a \nonumber \\
&& \delta A_\mu^a =0.
\label{omega}
\eea

By changing $Q \rightarrow Q-A$ in eq. (\ref{zaqcd}) we find
\bea
&& Z^{G_f}[A,J,\eta,{\bar \eta}]= e^{-i\int d^4x J \cdot A}~ \int [dQ] [d{\bar \psi}] [d \psi ] ~{\rm det}(\frac{\delta G_f^a(Q)}{\delta \beta^b}) \nonumber \\
&& e^{i\int d^4x [-\frac{1}{4}{F^a}_{\mu \nu}^2[Q] -\frac{1}{2 \alpha} (G_f^a(Q))^2+{\bar \psi} \dslash [Q] \psi + J \cdot Q + \eta {\bar \psi}
+{\bar \eta} \psi ]}
\label{zaqcd1}
\eea
where the gauge fixing term from eq. (\ref{ga}) becomes
\bea
G_f^a(Q) =\partial_\mu Q^{\mu a} + gf^{abc} A_\mu^b Q^{\mu c} - \partial_\mu A^{\mu a}=D_\mu[A] Q^{\mu a} - \partial_\mu A^{\mu a},
\label{gfa}
\eea
and eq. (\ref{omega}) becomes
\bea
\delta Q_\mu^a = -gf^{abc}\beta^b Q_\mu^c + \partial_\mu \beta^a.
\label{theta}
\eea
Note that eq. (\ref{theta}) is precisely the gauge transformation in QCD without the background field. Hence we
can anticipate a relation between QCD and the background field method of QCD when we relate $\beta^a(x)$ to the the
background field $A_\mu^a(x)$. For a pure gauge, $\beta^a(x)$ is related to the background field $A_\mu^a(x)$ in
eq. (\ref{gtqcd}).

In QCD (without the background field) the covariant gauge fixing term is
\bea
G_n^a(Q)=\partial_\mu Q^{\mu a}.
\label{gn}
\eea
We write this as
\bea
G_n^a(Q)=G_f^a(Q) +\Delta G_f^a(Q),~~~~~~~~~~~~~~~~~~\Delta G_f^a(Q) = -gf^{abc} A_\mu^b Q^{\mu c} + \partial_\mu A^{\mu a}.
\label{gngf}
\eea
Using this in eq. (\ref{zaqcd1}) we find
\bea
&& Z^{G_f+\Delta G_f}[A,J,\eta,{\bar \eta}]=e^{-i\int d^4x J \cdot A}~\int [dQ] [d{\bar \psi}] [d \psi ] ~{\rm det}(\frac{\delta (G_f^a(Q)+ \Delta G_f^a(Q))}{\delta \beta^b}) \nonumber \\
&& e^{i\int d^4x [-\frac{1}{4}{F^a}_{\mu \nu}^2[Q] -\frac{1}{2 \alpha} (G_f^a(Q) + \Delta G_f^a(Q))^2+{\bar \psi} \dslash [Q] \psi + J \cdot Q +{\bar \eta} \psi + \eta {\bar \psi}]}.
\label{zaqcd1a}
\eea
The ghost determinant becomes
\bea
&& {\rm det}(\frac{\delta (G_f^a(Q)+ \Delta G_f^a(Q))}{\delta \beta^b}) = {\rm det}(\frac{\delta G_f^a(Q)}{\delta \beta^b}) [1 + {\rm tr} [(\frac{\delta \Delta G_f^c(Q)}{\delta \beta^d})(\frac{\delta \beta^d}{\delta G_f^e(Q)})]+...] \nonumber \\
&& = {\rm det}(\frac{\delta G_f^a(Q)}{\delta \beta^b}) [1 +  (\frac{\delta \Delta G_f^c(Q)}{\delta \beta^d} ~\frac{\delta \beta^d}{\delta G_f^c(Q)})+...].
\label{dtgf}
\eea
Using eqs. (\ref{dtgf}) and (\ref{zaqcd1}) in (\ref{zaqcd1a}) we find
\bea
&& Z^{G_f+\Delta G_f}[A,J,\eta,{\bar \eta}]= Z^{G_f}[A,J,\eta,{\bar \eta}] \nonumber \\
&& +e^{-i\int d^4x J \cdot A}~\int [dQ] [d{\bar \psi}] [d \psi ]~{\rm det}(\frac{\delta G_f^a(Q)}{\delta \beta^b})
~e^{i\int d^4x [-\frac{1}{4}{F^a}_{\mu \nu}^2[Q] -\frac{1}{2 \alpha} (G_f^a(Q))^2+{\bar \psi} \dslash [Q] \psi + J \cdot Q +{\bar \eta} \psi
+ \eta {\bar \psi} ]} \nonumber \\
&& ~[i \int d^4x [ -\frac{1}{2 \alpha} (\Delta G_f^c(Q))^2-\frac{1}{\alpha} G_f^c(Q) \Delta G_f^c(Q)] + (\frac{\delta \Delta G_f^c(Q)}{\delta \beta^d} ~\frac{\delta \beta^d}{\delta G_f^c(Q)})+...].
\label{zaqcd2}
\eea

For infinitesimal gauge transformation we find from eq. (\ref{theta})
\bea
{Q'_\mu}^a = Q_\mu^a-gf^{abc}\beta^b Q_\mu^c + \partial_\mu \beta^a  = Q_\mu^a + \delta Q_\mu^a
\label{psieta1}
\eea
Changing the variables of integration from unprimed to primed variables in eq. (\ref{zaqcd1}) we find
\bea
&& Z^{G_f}[A,J,\eta,{\bar \eta}]=e^{-i\int d^4x J \cdot A}~\int [dQ'] [d{\bar \psi}'] [d \psi' ] \nonumber \\
&& ~{\rm det}(\frac{\delta G_f^a(Q')}{\delta \beta^b})
e^{i\int d^4x [-\frac{1}{4}{F^a}_{\mu \nu}^2[Q'] -\frac{1}{2 \alpha} (G_f^a(Q'))^2+{\bar \psi}' \dslash [Q'] \psi' + J \cdot Q' +{\bar \eta} \psi'
+ \eta {\bar \psi}' ]}.
\label{zaqcd1b}
\eea
This is because a change of variables from unprimed to primed variables does not change the value of the
integration. Under the infinitesimal gauge transformation, using eq. (\ref{psieta1}), we find \cite{muta}
\bea
&& [dQ'] =[dQ] ~{\rm det} [\frac{\partial {Q'}^a}{\partial Q^b}] = [dQ] ~{\rm det} [\delta^{ab} -gf^{abc} \beta^c]  \nonumber \\
&& =[dQ] [1- {\rm tr} gf^{abc} \beta^c + {\cal O}(\beta^2)+...] = [dQ] [1+ {\cal O}(\beta^2)+...]= [dQ],
\label{dq}
\eea
where we have made use of eq. (\ref{b1}).
Similarly under this infinitesimal gauge transformation the fermion fields transform accordingly {\it i.e.}
\bea
\psi' =\psi +\delta \psi,~~~~~~~~~~~~~~~~~{\bar \psi}'={\bar \psi}+\delta {\bar \psi}.
\label{pps}
\eea
Using eqs. (\ref{psieta1}) and (\ref{pps}) we find
\bea
[d{\bar \psi}'] [d \psi' ]=[d{\bar \psi}] [d \psi ],~~~~~~~~~~~{F^a}_{\mu \nu}^2[Q']={F^a}_{\mu \nu}^2[Q],~~~~~~~~~~~{\bar \psi}' \dslash [Q'] \psi'={\bar \psi} \dslash [Q] \psi
\label{psa}
\eea
which are gauge invariant.

Using eqs. (\ref{dq}) and (\ref{psa}) in eq. (\ref{zaqcd1b}) we find
\bea
&& Z^{G_f}[A,J,\eta,{\bar \eta}]=e^{-i\int d^4x J \cdot A}~\int [dQ] [d{\bar \psi}] [d \psi ] \nonumber \\
 && ~{\rm det}(\frac{\delta G_f^a(Q')}{\delta \beta^b})
~e^{i\int d^4x [-\frac{1}{4}{F^a}_{\mu \nu}^2[Q] -\frac{1}{2 \alpha} (G_f^a(Q'))^2+{\bar \psi} \dslash [Q] \psi + J \cdot Q' +{\bar \eta} \psi'
+ \eta {\bar \psi}' ]}.
\label{zaqcd1c}
\eea

Let us consider the background field to be a pure gauge as given by eq. (\ref{gtqcd}). For infinitesimal
gauge transformation parameter $\beta$, see eq. (\ref{b1}), we find from eq. (\ref{gtqcd})
\bea
A_\mu^a(x) = \partial_\mu \beta^a(x).
\label{am}
\eea
From eq. (\ref{gfa}) we find
\bea
G_f^a(Q') =\partial_\mu Q^{' \mu a} + gf^{abc} A_\mu^b Q^{' \mu c} - \partial_\mu A^{\mu a}.
\label{gfap}
\eea
which gives (by using eqs. (\ref{psieta1}) and (\ref{am}))
\bea
&& G_f^a(Q') =\partial_\mu [ Q^{\mu a}-gf^{abc}\beta^b Q^{\mu c} + \partial^\mu \beta^a]+ gf^{abc} A_\mu^b [Q^{\mu^c}-gf^{cde}\beta^d Q^{\mu e} + \partial^\mu \beta^c] - \partial_\mu A^{\mu a} \nonumber \\
&& = \partial_\mu Q^{\mu a}-gf^{abc}(\partial_\mu \beta^b) Q^{\mu c} -gf^{abc}\beta^b (\partial_\mu Q^{\mu c}) + \partial_\mu \partial^\mu \beta^a+ gf^{abc} A_\mu^b Q^{\mu c}-gf^{abc} A_\mu^bgf^{cde}\beta^d Q^{\mu e} \nonumber \\
&& + gf^{abc} A_\mu^b \partial^\mu \beta^c - \partial_\mu A^{\mu a} \nonumber \\
&& = \partial_\mu Q^{\mu a}-gf^{abc}A_\mu^b Q^{\mu c} -gf^{abc}\beta^b (\partial_\mu Q^{\mu c}) + \partial_\mu A^{\mu a}+ gf^{abc} A_\mu^b Q^{\mu c}-gf^{abc} A_\mu^bgf^{cde}\beta^d Q^{\mu e} \nonumber \\
&& + gf^{abc} A_\mu^b A^{\mu c} - \partial_\mu A^{\mu a} \nonumber \\
&& = \partial_\mu Q^{\mu a} -gf^{abc}\beta^b (\partial_\mu Q^{\mu c}) -gf^{abc} A_\mu^bgf^{cde}\beta^d Q^{\mu e} \nonumber \\
&& = \partial_\mu Q^{\mu a} -gf^{abc}\beta^b (\partial_\mu Q^{\mu c}) + {\cal O}(\beta^2).
\label{gfap1}
\eea
Note that $A$ is proportional to $\beta$, see eq. (\ref{am}). Hence
under an infinitesimal gauge
transformation (using eq. (\ref{psieta1})) we find from eq. (\ref{gfap1})
\bea
G_f^a(Q') =\partial_\mu Q^{\mu a} -gf^{abc}\beta^b (\partial_\mu Q^{\mu c})
\label{gqp}
\eea
where we have kept terms up to order ${\cal O}(\beta)$ and have neglected terms of order
${\cal O}(\beta^2)$, see eq. (\ref{b1}). Eq. (\ref{gqp}) gives
\bea
(G_f^a(Q'))^2 =(\partial_\mu Q^{\mu a})^2 -2gf^{abc}(\partial_\nu Q^{\nu a})\beta^b (\partial_\mu Q^{\mu c})
=(\partial_\mu Q^{\mu a})^2,
\label{gqp1}
\eea
where we have kept terms up to order ${\cal O}(\beta)$ and have neglected terms of order
${\cal O}(\beta^2)$ using eq. (\ref{b1}). Hence we find
\bea
(G_f^a(Q'))^2 =(\partial_\mu Q^{\mu a})^2=(G_n^a(Q))^2=(G_f^a(Q)+ \Delta G_f^a(Q))^2
\label{gqp3}
\eea
where we have used eqs. (\ref{gn}) and (\ref{gngf}). Using eq. (\ref{gqp}) we find
\bea
{\rm det} \frac{\delta G_f^a(Q')}{\delta \beta^b} ={\rm det}
\frac{ \delta [\partial_\mu Q^{\mu a} -gf^{abc}\beta^b (\partial_\mu Q^{\mu c})]}{\delta \beta^b}
={\rm det} [\frac{\delta  \partial_\mu Q^{\mu a}}{\delta \beta^b} -\frac{\delta [gf^{ab'c'}\beta^{b'} (\partial_\mu Q^{\mu c'})]}{\delta \beta^b}].
\label{gqp4}
\eea
Since from eq. (\ref{theta})
\bea
\delta Q_\mu^a \propto \beta^a
\label{theta1}
\eea
we find
\bea
\delta [gf^{abc}\beta^b (\partial_\mu Q^{\mu c})] \propto (\beta^a)^2.
\label{theta2}
\eea
Keeping terms up to order ${\cal O}(\beta)$ and neglecting terms of order ${\cal O}(\beta^2)$, see eq. (\ref{b1}),
we find from eq. (\ref{gqp4})
\bea
&& {\rm det} \frac{\delta G_f^a(Q')}{\delta \beta^b} =
{\rm det} \frac{\delta  \partial_\mu Q^{\mu a}}{\delta \beta^b} ={\rm det} \frac{\delta  [G_f^a(Q) + \Delta G_f^a(Q)]}{\delta \beta^b} \nonumber \\
&& ={\rm det} \frac{\delta  G_f^a(Q)}{\delta \beta^b} [1+\frac{\delta \Delta G_f^c(Q)}{\delta \beta^d}~\frac{\delta \beta^d}{\delta G_f^c(Q)}+...]
\label{gqp5}
\eea
where we have used eqs. (\ref{gn}) and (\ref{gngf}).
Using eqs. (\ref{gqp3}), (\ref{gqp5}), (\ref{psieta1}) and (\ref{pps}) in eq. (\ref{zaqcd1c}) we find
\bea
&& Z^{G_f}[A,J,\eta,{\bar \eta}]=e^{-i\int d^4x J \cdot A}~\int [dQ] [d{\bar \psi}] [d \psi ] ~{\rm det}(\frac{\delta G_f^a(Q)}{\delta \beta^b}) \nonumber \\
&& e^{i\int d^4x [-\frac{1}{4}{F^a}_{\mu \nu}^2[Q] -\frac{1}{2 \alpha} (G_f^a(Q))^2+{\bar \psi} \dslash [Q] \psi + J \cdot Q +{\bar \eta} \psi + \eta {\bar \psi}]}
~[1+ i\int d^4x [-\frac{1}{\alpha}G_f^c(Q) \Delta G_f^c(Q)-\frac{1}{2 \alpha}(\Delta G_f^c(Q))^2 \nonumber \\
&& +J \cdot \delta Q + {\bar \eta} \delta \psi + \eta \delta {\bar \psi}]+\frac{\delta \Delta G_f^c(Q)}{\delta \beta^d}~\frac{\delta \beta^d}{\delta G_f^c(Q)}+ ...].
\label{zaqcd1f}
\eea
Using eq. (\ref{zaqcd1}) in (\ref{zaqcd1f}) we find
\bea
&& Z^{G_f}[A,J,\eta,{\bar \eta}]= Z^{G_f}[A,~J,\eta,{\bar \eta}]\nonumber \\
&& + e^{-i\int d^4x J \cdot A}~\int [dQ] [d{\bar \psi}] [d \psi ] ~{\rm det}(\frac{\delta G_f^a(Q)}{\delta \beta^b})
~e^{i\int d^4x [-\frac{1}{4}{F^a}_{\mu \nu}^2[Q] -\frac{1}{2 \alpha} (G_f^a(Q))^2+{\bar \psi} \dslash [Q] \psi + J \cdot Q +{\bar \eta} \psi + \eta {\bar \psi}]} \nonumber \\
&& ~[i\int d^4x
[-\frac{1}{ \alpha}G_f^a(Q) \Delta G_f^a(Q)-\frac{1}{2 \alpha}(\Delta G_f^a(Q))^2 +J \cdot \delta Q + {\bar \eta} \delta \psi + \eta \delta
{\bar \psi}]
+\frac{\delta \Delta G_f^c(Q)}{\delta \beta^d}~\frac{\delta \beta^d}{\delta G_f^c(Q)}+...]. \nonumber \\
\label{zaqcd1g}
\eea
From the above equation we obtain our required identity
\bea
&& ~e^{-i\int d^4x J \cdot A}~\int [dQ] [d{\bar \psi}] [d \psi ] ~{\rm det}(\frac{\delta G_f^a(Q)}{\delta \beta^b})
~e^{i\int d^4x [-\frac{1}{4}{F^a}_{\mu \nu}^2[Q] -\frac{1}{2 \alpha} (G_f^a(Q))^2+{\bar \psi} \dslash [Q] \psi + J \cdot Q +{\bar \eta} \psi + \eta  {\bar \psi}]} \nonumber \\
&& ~[i\int d^4x [-\frac{1}{ \alpha}G_f^c(Q) \Delta G_f^c(Q)-\frac{1}{2 \alpha}(\Delta G_f^c(Q))^2 +J \cdot \delta Q
+ {\bar \eta} \delta \psi + \eta \delta {\bar \psi} ] \nonumber \\
&& +\frac{\delta \Delta G_f^c(Q)}{\delta \beta^d}~\frac{\delta \beta^d}{\delta G_f^c(Q)}+...]=0.
\label{zaqcd1h}
\eea
Eq. (\ref{zaqcd1h}) is the gauge fixing identity by varying the gauge fixing term in $Z[A,J,\eta,{\bar \eta}]$ in
the background field method of QCD in pure gauge.
\section{ Derivation of Gauge Fixing Identity by Using General Non-Covariant Gauge Fixing }
The generating functional in the background field method of QCD with general non-covariant gauge fixing is given by
\bea
&& Z^{G}[A,J,\eta,{\bar \eta}]=\int [dQ] [d{\bar \psi}] [d \psi ] ~{\rm det}(\frac{\delta G^a(Q)}{\delta \omega^b}) \nonumber \\
&& e^{i\int d^4x [-\frac{1}{4}{F^a}_{\mu \nu}^2[A+Q] -\frac{1}{2 \alpha}
(G^a(Q))^2+{\bar \psi} \dslash [A+Q] \psi + J \cdot Q +{\bar \eta} \psi + \eta {\bar \psi} ]}
\label{zaqcdn}
\eea
where
\bea
G^a(Q) =\frac{\eta^\mu \eta^\nu}{\eta^2} ~(\partial_\mu Q_\nu^a + gf^{abc} A_\mu^b Q_\nu^c)=\frac{\eta^\mu \eta^\nu}{\eta^2}~D_\mu[A]Q_\nu^a
\label{gan}
\eea
is the gauge fixing term in general non-covariant gauges \cite{noncov,noncov1} with $\eta^\mu$ being an arbitrary but constant four vector.
By changing $Q_\mu^a \rightarrow Q_\mu^a -A_\mu^a$ in eq. (\ref{zaqcdn}) we find
\bea
&& Z^{G_f}[A,J,\eta,{\bar \eta}]= e^{-i\int d^4x J \cdot A}~ \int [dQ] [d{\bar \psi}] [d \psi ] ~{\rm det}(\frac{\delta G_f^a(Q)}{\delta \beta^b}) \nonumber \\
&& e^{i\int d^4x [-\frac{1}{4}{F^a}_{\mu \nu}^2[Q] -\frac{1}{2 \alpha} (G_f^a(Q))^2+{\bar \psi} \dslash [Q] \psi + J \cdot Q + \eta {\bar \psi}
+{\bar \eta} \psi ]}
\label{zaqcd1n}
\eea
where
\bea
&& G_f^a(Q) =\frac{\eta^\mu \eta^\nu}{\eta^2}~(\partial_\mu Q_\nu^a + gf^{abc} A_\mu^b Q_\nu^c - \partial_\mu A_\nu^a)-\frac{1}{\eta^2}~gf^{abc} (\eta \cdot A^b) (\eta \cdot A^c) \nonumber \\
&& =\frac{\eta^\mu \eta^\nu}{\eta^2}~(D_\mu[A] Q_\nu^a) - \frac{\eta^\mu \eta_\nu}{\eta^2}~\partial_\mu A_\nu^a.
\label{gfan}
\eea
In QCD (without the background field) the generating functional with general non-covariant gauge fixing is given by
\bea
&& Z[J,\eta,{\bar \eta}]=\int [dQ] [d{\bar \psi}] [d \psi ] ~{\rm det}(\frac{\delta G_n^a(Q)}{\delta \beta^b}) \nonumber \\
&& e^{i\int d^4x [-\frac{1}{4}{F^a}_{\mu \nu}^2[Q] -\frac{1}{2 \alpha} (G_n^a(Q))^2+{\bar \psi} \dslash [Q] \psi + J \cdot Q + \eta {\bar \psi}
+{\bar \eta} \psi ]}
\label{zqcdn}
\eea
where
\bea
G_n^a(Q)=\frac{\eta^\mu \eta^\nu}{\eta^2}~\partial_\mu Q_\nu^a
\label{gnn}
\eea
is the gauge fixing term in general non-covariant gauges \cite{noncov,noncov1}. We write this as
\bea
G_n^a(Q)=G_f^a(Q) +\Delta G_f^a(Q),~~~~~~~~~~~~~~~~~~\Delta G_f^a(Q) =\frac{\eta^\mu \eta^\nu}{\eta^2}~( -gf^{abc} A_\mu^b Q_\nu^c + \partial_\mu A_\nu^a).
\label{gngfn}
\eea

Changing the variables of integration from unprimed to primed variables in eq. (\ref{zaqcd1n}) we find
\bea
&& Z^{G_f}[A,J,\eta,{\bar \eta}]=e^{-i\int d^4x J \cdot A}~\int [dQ] [d{\bar \psi}] [d \psi ] \nonumber \\
 && ~{\rm det}(\frac{\delta G_f^a(Q')}{\delta \beta^b})
~e^{i\int d^4x [-\frac{1}{4}{F^a}_{\mu \nu}^2[Q] -\frac{1}{2 \alpha} (G_f^a(Q'))^2+{\bar \psi} \dslash [Q] \psi + J \cdot Q' +{\bar \eta} \psi'
+ \eta {\bar \psi}' ]}
\label{zaqcd1cn}
\eea
which is similar to eq. (\ref{zaqcd1c}) but with $G_f^a(Q)$ given by eq. (\ref{gfan}).

By using eqs. (\ref{psieta1}) and (\ref{am}) we find from eq. (\ref{gfan})
\bea
&& G_f^a(Q') = \frac{\eta^\mu \eta_\nu}{\eta^2}~(\partial_\mu [ Q^{\nu a}-gf^{abc}\beta^b Q^{\nu c} + \partial^\nu \beta^a]+ gf^{abc} A_\mu^b [Q^{\nu c}-gf^{cde}\beta^d Q^{\nu e} + \partial^\nu \beta^c] - \partial_\mu A^{\nu a}) \nonumber \\
&& = \frac{\eta^\mu \eta_\nu}{\eta^2}~(\partial_\mu Q^{\nu a}-gf^{abc}(\partial_\mu \beta^b) Q^{\nu c} -gf^{abc}\beta^b (\partial_\mu Q^{\nu c}) + \partial_\mu \partial^\nu \beta^a+ gf^{abc} A_\mu^b Q^{\nu c}-gf^{abc} A_\mu^bgf^{cde}\beta^d Q^{\nu e} \nonumber \\
&& + gf^{abc} A_\mu^b \partial^\nu \beta^c - \partial_\mu A^{\nu a}) \nonumber \\
&& = \frac{\eta^\mu \eta_\nu}{\eta^2}~(\partial_\mu Q^{\nu a}-gf^{abc}A_\mu^b Q^{\nu c} -gf^{abc}\beta^b (\partial_\mu Q^{\nu c}) + \partial_\mu A^{\nu a}+ gf^{abc} A_\mu^b Q^{\nu c}-gf^{abc} A_\mu^bgf^{cde}\beta^d Q^{\nu e} \nonumber \\
&& + gf^{abc} A_\mu^b A^{\nu c} - \partial_\mu A^{\nu a}) \nonumber \\
&& = \frac{\eta^\mu \eta_\nu}{\eta^2}~(\partial_\mu Q^{\nu a} -gf^{abc}\beta^b (\partial_\mu Q^{\nu c}) -gf^{abc} A_\mu^bgf^{cde}\beta^d Q^{\nu e}) +\frac{1}{\eta^2}~gf^{abc} (\eta \cdot A^b)(\eta \cdot A^c) \nonumber \\
&& = \frac{\eta^\mu \eta_\nu}{\eta^2}~(\partial_\mu Q^{\nu a} -gf^{abc}\beta^b (\partial_\mu Q^{\nu c}) -gf^{abc} A_\mu^bgf^{cde}\beta^d Q^{\nu e}) \nonumber \\
&& = \frac{\eta^\mu \eta_\nu}{\eta^2}~(\partial_\mu Q^{\nu a} -gf^{abc}\beta^b (\partial_\mu Q^{\nu c})) + {\cal O}(\beta^2).
\label{gfap1n}
\eea
Note that $A$ is proportional to $\beta$, see eq. (\ref{am}). Hence under an infinitesimal gauge
transformation (using eq. (\ref{psieta1})) we find from eq. (\ref{gfap1n})
\bea
G_f^a(Q') =\frac{\eta^\mu \eta_\nu}{\eta^2}~(\partial_\mu Q^{\nu a} -gf^{abc}\beta^b (\partial_\mu Q^{\nu c}))
\label{gqpn}
\eea
where we have kept terms up to order ${\cal O}(\beta)$ and have neglected terms of order
${\cal O}(\beta^2)$, see eq. (\ref{b1}). Eq. (\ref{gqpn}) gives
\bea
(G_f^a(Q'))^2 =(\frac{\eta^\mu \eta_\nu}{\eta^2}~\partial_\mu Q^{\nu a})^2
-2gf^{abc}\beta^b \frac{1}{\eta^4}(\eta \cdot \partial \eta \cdot Q^{c})~ (\eta \cdot \partial \eta \cdot Q^{a})
+{\cal O} (\beta^2).
\label{gqp1n}
\eea
Keeping terms up to order ${\cal O}(\beta)$ and neglecting terms of order ${\cal O}(\beta^2)$, see eq. (\ref{b1}),
we find from eq. (\ref{gqp1n})
\bea
(G_f^a(Q'))^2 =(\frac{\eta^\mu \eta_\nu}{\eta^2}~\partial_\mu Q^{\nu a})^2=(G_n^a(Q))^2=(G_f^a(Q)+ \Delta G_f^a(Q))^2
\label{gqp3n}
\eea
where we have used eqs. (\ref{gnn}) and (\ref{gngfn}). Using eq. (\ref{gqpn}) we find
\bea
{\rm det} \frac{\delta G_f^a(Q')}{\delta \beta^b} ={\rm det}
\frac{\frac{\eta^\mu \eta_\nu}{\eta^2}~( \delta [\partial_\mu Q^{\nu a} -gf^{abc}\beta^b (\partial_\mu Q^{\nu c})])}{\delta \beta^b}
={\rm det} [\frac{\eta^\mu \eta_\nu}{\eta^2}~(\frac{\delta  \partial_\mu Q^{\nu a}}{\delta \beta^b} -\frac{\delta [gf^{ab'c'}\beta^{b'} (\partial_\mu Q^{\nu c'})]}{\delta \beta^b})]. \nonumber \\
\label{gqp4n}
\eea
Keeping terms up to order ${\cal O}(\beta)$ and neglecting terms of order ${\cal O}(\beta^2)$, see eqs. (\ref{b1}) and (\ref{theta2}),
we find from eq. (\ref{gqp4n})
\bea
&& {\rm det} \frac{\delta G_f^a(Q')}{\delta \beta^b} =
{\rm det} \frac{\frac{\eta^\mu \eta_\nu}{\eta^2}~(\delta  \partial_\mu Q^{\nu a})}{\delta \beta^b} ={\rm det} \frac{\delta  [G_f^a(Q) + \Delta G_f^a(Q)]}{\delta \beta^b} \nonumber \\
&& ={\rm det} \frac{\delta  G_f^a(Q)}{\delta \beta^b} [1+\frac{\delta \Delta G_f^c(Q)}{\delta \beta^d}~\frac{\delta \beta^d}{\delta G_f^c(Q)}+...]
\label{gqp5n}
\eea
where we have used eqs. (\ref{gnn}) and (\ref{gngfn}).
Using eqs. (\ref{gqp3n}), (\ref{gqp5n}), (\ref{psieta1}) and (\ref{pps}) in eq. (\ref{zaqcd1cn}) we find
\bea
&& Z^{G_f}[A,J,\eta,{\bar \eta}]=e^{-i\int d^4x J \cdot A}~\int [dQ] [d{\bar \psi}] [d \psi ] ~{\rm det}(\frac{\delta G_f^a(Q)}{\delta \beta^b}) \nonumber \\
&& e^{i\int d^4x [-\frac{1}{4}{F^a}_{\mu \nu}^2[Q] -\frac{1}{2 \alpha} (G_f^a(Q))^2+{\bar \psi} \dslash [Q] \psi + J \cdot Q +{\bar \eta} \psi + \eta {\bar \psi}]}
~[1+ i\int d^4x [-\frac{1}{\alpha}G_f^c(Q) \Delta G_f^c(Q)-\frac{1}{2 \alpha}(\Delta G_f^c(Q))^2 \nonumber \\
&& +J \cdot \delta Q + {\bar \eta} \delta \psi + \eta \delta {\bar \psi}]+\frac{\delta \Delta G_f^c(Q)}{\delta \beta^d}~\frac{\delta \beta^d}{\delta G_f^c(Q)}+ ...]
\label{zaqcd1fn}
\eea
where $G_f^a(Q)$ is given by eq. (\ref{gfan}). Using eq. (\ref{zaqcdn}) we find from eq. (\ref{zaqcd1fn})
\bea
&& Z^{G_f}[A,J,\eta,{\bar \eta}]= Z^{G_f}[A,~J,\eta,{\bar \eta}]\nonumber \\
&& + e^{-i\int d^4x J \cdot A}~\int [dQ] [d{\bar \psi}] [d \psi ] ~{\rm det}(\frac{\delta G_f^a(Q)}{\delta \beta^b})
~e^{i\int d^4x [-\frac{1}{4}{F^a}_{\mu \nu}^2[Q] -\frac{1}{2 \alpha} (G_f^a(Q))^2+{\bar \psi} \dslash [Q] \psi + J \cdot Q +{\bar \eta} \psi + \eta {\bar \psi}]} \nonumber \\
&& ~[i\int d^4x
[-\frac{1}{ \alpha}G_f^a(Q) \Delta G_f^a(Q)-\frac{1}{2 \alpha}(\Delta G_f^a(Q))^2 +J \cdot \delta Q + {\bar \eta} \delta \psi + \eta \delta
{\bar \psi}]
+\frac{\delta \Delta G_f^c(Q)}{\delta \beta^d}~\frac{\delta \beta^d}{\delta G_f^c(Q)}+...]. \nonumber \\
\label{zaqcd1gn}
\eea
From the above equation we obtain our required gauge fixing identity
\bea
&& ~e^{-i\int d^4x J \cdot A}~\int [dQ] [d{\bar \psi}] [d \psi ] ~{\rm det}(\frac{\delta G_f^a(Q)}{\delta \beta^b})
~e^{i\int d^4x [-\frac{1}{4}{F^a}_{\mu \nu}^2[Q] -\frac{1}{2 \alpha} (G_f^a(Q))^2+{\bar \psi} \dslash [Q] \psi + J \cdot Q +{\bar \eta} \psi + \eta  {\bar \psi}]} \nonumber \\
&& ~[i\int d^4x [-\frac{1}{ \alpha}G_f^c(Q) \Delta G_f^c(Q)-\frac{1}{2 \alpha}(\Delta G_f^c(Q))^2 +J \cdot \delta Q
+ {\bar \eta} \delta \psi + \eta \delta {\bar \psi} ] \nonumber \\
&& +\frac{\delta \Delta G_f^c(Q)}{\delta \beta^d}~\frac{\delta \beta^d}{\delta G_f^c(Q)}+...]=0
\label{zaqcd1hn}
\eea
by using general non-covariant gauge fixing.

Eq. (\ref{zaqcd1hn}) is exactly same as eq. (\ref{zaqcd1h}) in covariant gauge fixing except that $G_f^a(Q)$ replaced
by the general non-covariant gauge fixing term given by eq. (\ref{gfan}). This completes the derivation of gauge fixing identity
by using the general non-covariant gauge fixing.
\section{ Derivation of Gauge Fixing Identity by Using General Coulomb Gauge Fixing }
The generating functional in the background field method of QCD with general Coulomb gauge fixing is given by
\bea
&& Z^{G}[A,J,\eta,{\bar \eta}]=\int [dQ] [d{\bar \psi}] [d \psi ] ~{\rm det}(\frac{\delta G^a(Q)}{\delta \omega^b}) \nonumber \\
&& e^{i\int d^4x [-\frac{1}{4}{F^a}_{\mu \nu}^2[A+Q] -\frac{1}{2 \alpha}
(G^a(Q))^2+{\bar \psi} \dslash [A+Q] \psi + J \cdot Q +{\bar \eta} \psi + \eta {\bar \psi} ]}
\label{zaqcdc}
\eea
where
\bea
G^a(Q) =[g^{\mu \nu}-\frac{n^\mu n^\nu}{n^2}] ~(\partial_\mu Q_\nu^a + gf^{abc} A_\mu^b Q_\nu^c)=
[g^{\mu \nu}-\frac{\eta^\mu n^\nu}{n^2}]~D_\mu[A]Q_\nu^a
\label{gac}
\eea
is the gauge fixing term in general Coulomb gauge \cite{noncov} with
\bea
n^\mu =(1,0,0,0).
\eea
By changing $Q_\mu^a \rightarrow Q_\mu^a -A_\mu^a$ in eq. (\ref{zaqcdc}) we find
\bea
&& Z^{G_f}[A,J,\eta,{\bar \eta}]= e^{-i\int d^4x J \cdot A}~ \int [dQ] [d{\bar \psi}] [d \psi ] ~{\rm det}(\frac{\delta G_f^a(Q)}{\delta \beta^b}) \nonumber \\
&& e^{i\int d^4x [-\frac{1}{4}{F^a}_{\mu \nu}^2[Q] -\frac{1}{2 \alpha} (G_f^a(Q))^2+{\bar \psi} \dslash [Q] \psi + J \cdot Q + \eta {\bar \psi}
+{\bar \eta} \psi ]}
\label{zaqcd1cc}
\eea
where
\bea
&& G_f^a(Q) =[g^{\mu \nu}-\frac{n^\mu n^\nu}{n^2}]~(\partial_\mu Q_\nu^a + gf^{abc} A_\mu^b Q_\nu^c - \partial_\mu A_\nu^a)-gf^{abc} A_\mu^b A^{\mu c}+\frac{1}{n^2}~gf^{abc} (n \cdot A^b) (n \cdot A^c) \nonumber \\
&& =[g^{\mu \nu}-\frac{n^\mu n^\nu}{n^2}]~(D_\mu[A] Q_\nu^a) - [g^{\mu \nu}-\frac{n^\mu n^\nu}{n^2}]~\partial_\mu A_\nu^a.
\label{gfac}
\eea
In QCD (without the background field) the generating functional with general Coulomb gauge fixing is given by
\bea
 Z[J,\eta,{\bar \eta}]=\int [dQ] [d{\bar \psi}] [d \psi ] ~{\rm det}(\frac{\delta G_n^a(Q)}{\delta \beta^b}) e^{i\int d^4x [-\frac{1}{4}{F^a}_{\mu \nu}^2[Q] -\frac{1}{2 \alpha} (G_n^a(Q))^2+{\bar \psi} \dslash [Q] \psi + J \cdot Q + \eta {\bar \psi}
+{\bar \eta} \psi ]}
\label{zqcdc}
\eea
where
\bea
G_n^a(Q)=[g^{\mu \nu}-\frac{n^\mu n^\nu}{n^2}]~\partial_\mu Q_\nu^a
\label{gnc}
\eea
is the gauge fixing term in general Coulomb gauge \cite{noncov}.

Hence by replacing $\frac{\eta^\mu \eta^\nu}{\eta^2} \rightarrow [g^{\mu \nu}-\frac{n^\mu n^\nu}{n^2}]$ everywhere in the derivations
in the previous section we find
\bea
&& ~e^{-i\int d^4x J \cdot A}~\int [dQ] [d{\bar \psi}] [d \psi ] ~{\rm det}(\frac{\delta G_f^a(Q)}{\delta \beta^b})
~e^{i\int d^4x [-\frac{1}{4}{F^a}_{\mu \nu}^2[Q] -\frac{1}{2 \alpha} (G_f^a(Q))^2+{\bar \psi} \dslash [Q] \psi + J \cdot Q +{\bar \eta} \psi + \eta  {\bar \psi}]} \nonumber \\
&& ~[i\int d^4x [-\frac{1}{ \alpha}G_f^c(Q) \Delta G_f^c(Q)-\frac{1}{2 \alpha}(\Delta G_f^c(Q))^2 +J \cdot \delta Q
+ {\bar \eta} \delta \psi + \eta \delta {\bar \psi} ] \nonumber \\
&& +\frac{\delta \Delta G_f^c(Q)}{\delta \beta^d}~\frac{\delta \beta^d}{\delta G_f^c(Q)}+...]=0
\label{zaqcd1hc}
\eea
by using general Coulomb gauge fixing.

Eq. (\ref{zaqcd1hc}) is exactly same as eqs. (\ref{zaqcd1h}) (or eq. (\ref{zaqcd1hn})) with covariant gauge fixing
(or general non-covariant gauge fixing) except that $G_f^a(Q)$ is replaced by general Coulomb gauge fixing term given by eq.
(\ref{gfac}) . This completes the derivation of gauge fixing identity by using the general Coulomb gauge fixing.
\section{ Relation Between $Z[J,\eta,{\bar \eta}]$ and $Z[A,J,\eta,{\bar \eta}]$ in Pure Gauge }
Using eq. (\ref{gn}) and (\ref{gngf}) in (\ref{zaqcd1}) we obtain
\bea
&& Z^{G_f+\Delta G_f}[A,J,\eta,{\bar \eta}]= e^{-i\int d^4x J \cdot A}~\int [dQ] [d{\bar \psi}] [d \psi ] ~{\rm det}(\frac{~\delta G_n^a(Q)}{\delta \beta^b}) \nonumber \\
&& e^{i\int d^4x [-\frac{1}{4}{F^a}_{\mu \nu}^2[Q] -\frac{1}{2 \alpha} ({G_n^a}(Q))^2+{\bar \psi} \dslash [Q] \psi + J \cdot Q +{\bar \eta} \psi + \eta  {\bar \psi} ]}~ =e^{-i\int d^4x J \cdot A}~\times~Z_{\rm QCD}[J,\eta,{\bar \eta}]
\label{zaqcd1nf}
\eea
where $Z_{\rm QCD}[J,\eta,{\bar \eta}]$ is the generating functional in QCD without the background field.
Subtracting eq. (\ref{zaqcd1h}) from eq. (\ref{zaqcd2}) we find
\bea
&& Z^{G_f+\Delta G_f}[A,J,\eta,{\bar \eta}]= Z^{G_f}[A,J,\eta,{\bar \eta}] -e^{-i\int d^4x J \cdot A}~\int [dQ] [d{\bar \psi}] [d \psi ]~{\rm det}(\frac{\delta G_f^a(Q)}{\delta \beta^b}) \nonumber \\
&&~e^{i\int d^4x [-\frac{1}{4}{F^a}_{\mu \nu}^2[Q] -\frac{1}{2 \alpha} (G_f^a(Q))^2+{\bar \psi} \dslash [Q] \psi + J \cdot Q +{\bar \eta} \psi
+ \eta {\bar \psi} ]} ~[i \int d^4x [ J \cdot \delta Q
+ {\bar \eta} \delta \psi + \eta \delta {\bar \psi}+...]].
\label{zaqcd2new}
\eea
From eq. (\ref{zaqcd1}) we denote
\bea
Z_{\rm background ~QCD}[A,J,\eta,{\bar \eta}]~~=~~Z^{G_f}[A,J,\eta,{\bar \eta}].
\label{bcqcd}
\eea

Hence from eqs. (\ref{zaqcd1}), (\ref{zaqcd2new}) and (\ref{bcqcd}) we find
\bea
&& Z_{\rm QCD}[J,\eta,{\bar \eta}] ~=e^{i\int d^4x J \cdot A}~\times~Z_{\rm background ~QCD}[A,J,\eta,{\bar \eta}]-\int [dQ] [d{\bar \psi}] [d \psi ]~{\rm det}(\frac{\delta G_f^a(Q)}{\delta \beta^b}) \nonumber \\
&&~e^{i\int d^4x [-\frac{1}{4}{F^a}_{\mu \nu}^2[Q] -\frac{1}{2 \alpha} (G_f^a(Q))^2+{\bar \psi} \dslash [Q] \psi + J \cdot Q +{\bar \eta} \psi
+ \eta {\bar \psi} ]} ~[i \int d^4x [ J \cdot \delta Q
+ {\bar \eta} \delta \psi + \eta \delta {\bar \psi}+...]],
\label{final}
\eea
where the covariant gauge fixing term $G_f^a(Q)$ is given by eq. (\ref{gfa}).

Similarly using eq. (\ref{zaqcd1hn}) with general non-covariant gauge fixing
we arrive at eq. (\ref{final}) with general non-covariant gauge fixing term
$G_f^a(Q)$ given by eq. (\ref{gfan}). Using eq. (\ref{zaqcd1hc}) with general
Coulomb gauge fixing we arrive at eq. (\ref{final}) with general Coulomb gauge
fixing term $G_f^a(Q)$ given by eq. (\ref{gfac}).

Hence we find that eq. (\ref{final}) is the relation between generating functional in QCD
and generating functional in the background field method of QCD in pure gauge with covariant,
general non-covariant and general Coulomb gauge fixings respectively. We have used eq. (\ref{final})
in \cite{nkv} to prove factorization of soft and collinear divergences at high energy colliders. We
have also used this equation in \cite{nkm} is to prove factorization of fragmentation function in
non-equilibrium QCD which can be experimentally applicable at RHIC and LHC \cite{nayakfr}.
\section{Conclusions}
Establishing a relation between $Z[A,J,\eta, {\bar \eta}]$ in the background field method of QCD
in pure gauge and $Z[J,\eta,{\bar \eta}]$ in QCD can be useful in studying properties of certain
non-perturbative physical quantities in QCD. However, unlike QED, establishing a relation between these two
is not easy because the gauge fixing term in QCD is different from background field
gauge fixing term in the background field method of QCD. For this purpose we have derived a gauge fixing identity
by varying the covariant gauge fixing term in $Z[A,J,\eta, {\bar \eta}]$ in the
background field method of QCD in pure gauge. We have made a detailed analysis of this gauge fixing identity
by using general non-covariant and general Coulomb gauge fixing terms. We have found that the gauge fixing identity
holds for covariant, general non-covariant and general Coulomb gauge fixings respectively.

Using this gauge fixing identity we have established a relation (given by eq. (\ref{final})) between
$Z[J,\eta,{\bar \eta}]$ in QCD and $Z[A,J,\eta, {\bar \eta}]$ in background field method of QCD in pure gauge. We have
used this gauge fixing identity in \cite{nkv} to prove factorization of soft and collinear divergences at high energy
colliders. We have also used this identity in \cite{nkm} to prove factorization of fragmentation function in non-equilibrium
QCD which can be experimentally applicable at RHIC and LHC \cite{nayakfr}.

\acknowledgements

I thank Peter van Nieuwenhuizen for useful discussions. This work was supported
in part by Department of Energy under contracts DE-FG02-91ER40664, DE-FG02-04ER41319
and DE-FG02-04ER41298.

\end{document}